\documentclass[%
 reprint,
 amsmath,amssymb,
 aps,
floatfix,
]{revtex4-1}
\usepackage{ulem}

\usepackage{soul,color}
\usepackage{xcolor}
\usepackage{amsmath}
\usepackage{graphicx}
\usepackage{dcolumn}
\usepackage{bm}
\graphicspath{{figures/}}

\begin{document}

\preprint{APS/123-QED}

\title{Interaction-Enhanced Group Velocity of Bosons in the Flat Band of an Optical Kagome Lattice}

\author{Tsz-Him Leung$^1$, Malte N. Schwarz$^{1,3}$, Shao-Wen Chang$^1$, Charles D. Brown$^1$, Govind Unnikrishnan$^4$, Dan Stamper-Kurn$^{1,2}$}

\affiliation{
    $^1$Department of Physics, University of California, Berkeley CA 94720 \\
    $^2$Materials Sciences Division, Lawrence Berkeley National Laboratory, Berkeley, CA 94720 \\
    $^3$Fakult{\"a}t f{\"u}r Physik und Astronomie, Universit{\"a}t W{\"u}rzburg, 97074 W{\"u}rzburg, Germany \\
    $^4$Institut f{\"u}r Experimentalphysik und Zentrum f{\"u}r Quantenphysik, Universit{\"a}t Innsbruck, 6020 Innsbruck, Austria}
\date{\today}

\begin{abstract}

Geometric frustration of particle motion in a kagome lattice causes the single-particle band structure to have a flat s-orbital band.  We probe this band structure by exciting a Bose-Einstein condensate into excited Bloch states of an optical kagome lattice, and then measuring the group velocity through the atomic momentum distribution.  We find that interactions renormalize the band structure of the kagome lattice, greatly increasing the dispersion of the third band that, according to non-interacting band theory, should be nearly non-dispersing.  Measurements at various lattice depths and gas densities agree quantitatively with predictions of the lattice Gross-Pitaevskii equation, indicating that the observed distortion of band structure is caused by the distortion of the overall lattice potential away from the kagome geometry by interactions.

\end{abstract}

\maketitle

Band structure describes the states of motion of non-interacting particles within a spatially periodic potential, and serves as a key ingredient for understanding properties of materials, the propagation of light in photonic crystals, and the transport of ultracold atoms within optical lattices.  
In some materials, interactions cause band structure to differ strongly from the non-interacting case, an effect known as band-structure renormalization.  Such interaction-driven renormalization can be particularly important in heavy-fermion materials, where the Fermi energy lies within a band with very small dispersion (a flat band) \cite{zwic92heavy}.  

Flat bands have also been realized in optical lattices.  Specifically, in the two-dimensional kagome \cite{jo12kag} and Lieb \cite{taie15lieb} lattices, geometric frustration of particle motion produces non-dispersing bands.  In the tight-binding limit, with the tunneling energy between neighboring sites $i$ and $j$ defined as $-J (\hat{a}^\dagger_i \hat{a}_j + \hat{a}^\dagger_j \hat{a}_i)$,  a flat band emerges as the third and second  bands of the $J>0$ kagome and Lieb lattices, respectively.  Here, $\hat{a}_i$ ($\hat{a}_i^\dagger$) is  the bosonic particle annihilation (creation) operator at site $i$.

Systems of interacting bosons or fermions  that equilibrate within flat bands are the subject of intense theoretical interest \cite{hube10flat, you12kagome, maiti19fermi, hui17soc, tovmasyan13pair,tasa92,imad00,iglo14}.  Specifically, for interacting bosons equilibrating in the flat ground band of a $J<0$ kagome lattice, You et al.\ \cite{you12kagome} propose that interactions renormalize the band structure, causing a stable superfluid to form at the $K$ or $\Gamma$ points of the Brilliouin zone, where, self-consistently, the band energy is  minimized.

In this work, we probe the effects of interactions on the flat band of an optical kagome lattice with a gas of bosons.  A Bose-Einstein condensate of $^{87}$Rb atoms is prepared at rest, accelerated, and then loaded adiabatically into an excited Bloch state of the ($J>0$) kagome lattice with variable quasimomentum $\mathbf{q}$ and band index $n$.  We characterize this far-from-equilibrium state by measuring its momentum distribution and group velocity $\mathbf{v}_g$.  
We find the group velocity  for atoms in the  $n=3$ band of the kagome lattice to be significantly larger than expected for non-interacting atoms. Through experiments and numerical calculations, we 
confirm that the un-flattening of the $n=3$ kagome band results from interaction-driven band-structure renormalization.  Our work verifies the physical picture suggested by Ref.\ \cite{you12kagome}, and, more generally, demonstrates that the transport properties of lattice-trapped atoms can be significantly influenced by interactions. Recently, the distortion of the flat band of the Lieb lattice by an interacting gas in a superposition of band states has also been observed  \cite{ozaw17interaction}.

\begin{figure}[t]
\includegraphics{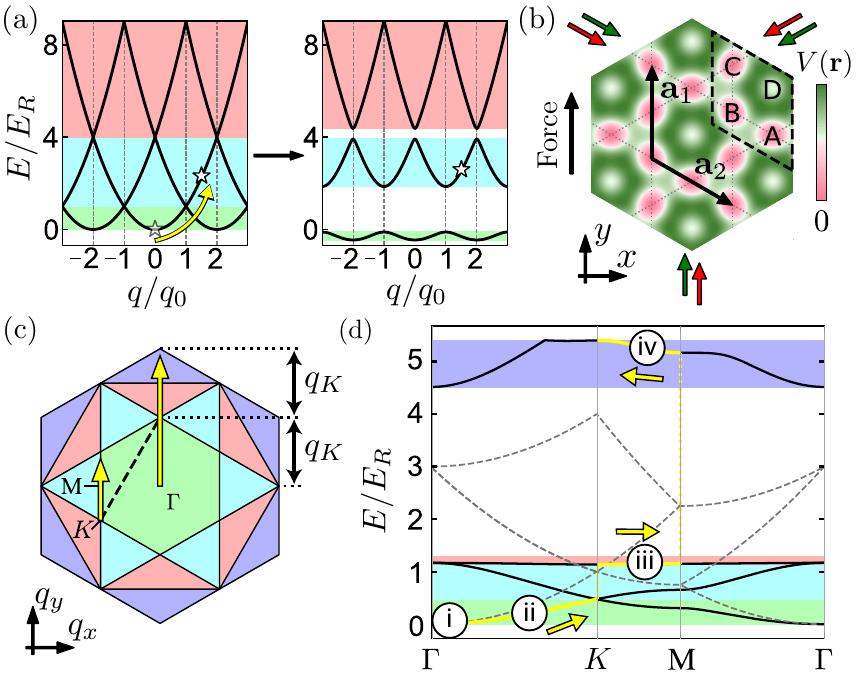}
\caption{Experimental scheme. (a) The band structure of a 1D lattice drawn in the extended zone scheme at zero (left) and non-zero (right) lattice depth.  A particle accelerated to the $n$-th Brillouin zone ($ n-1< |k|/q_0 < n $) is loaded into the $n$-th band as the lattice is ramped up adiabatically. $q_0$ is the magnitude of the basis reciprocal lattice vector and $E_R$ is the lattice recoil energy. (b) The optical kagome lattice is constructed by overlaying triangular lattices using short- (532 nm, green) and long-wavelength (1064, shown in red) light. Primitive lattice vectors $\mathbf{a}_{1,2}$ are shown. The four sites in a unit cell are labelled by letters A-D. 
(c) The first four Brillouin zones of the kagome lattice.  Acceleration of the condensate along $\mathbf{y}$ maps atoms sequentially into the $n=1$ ($\Gamma$ to $K$), $n=3$ ($K$ to $M$), and then $n=4$ ($M$ to $K$) bands.  Displacement by a reciprocal lattice vector (thick dashed line) reveals the path within the $n=3,4$ zones in the reduced zone scheme.  (d) Non-interacting band structure of the optical kagome lattice. Solid black lines: (V$_{\text{SW}}$, V$_{\text{LW}}$) = $h \times$ (25, 15) kHz.  Dashed lines: zero lattice depth.  The yellow trajectory indicates the Bloch states to which atoms accelerated along the trajectory in (c) are adiabatically connected. The circled labels indicate where the data in Fig. \ref{fig:GV}(b) are taken. Here $E_R = 2.0$ kHz is the kagome-lattice recoil energy.}
\label{fig:scheme}
\end{figure}

Band theory provides the eigenstates of single particles in a spatially periodic potential as Bloch states $\Psi^{(n)}_\mathbf{q}(\mathbf{r})$ with energies $E_n(\mathbf{q})$. In the limit of vanishing potential depth, the band structure of any lattice approaches the dispersion relation of a free particle with momentum $\mathbf{p} = \hbar \mathbf{k}$, and Bloch states map onto plane waves $\Psi^{(n)}_\mathbf{q}(\mathbf{r}) \sim \exp(i \mathbf{k} \cdot \mathbf{r})$, where $\mathbf{k}$ lies in the $n$-th Brillouin zone and $\mathbf{k} = \mathbf{q}$ modulo reciprocal lattice vectors.  This mapping provides a three-step protocol to transport all atoms within a Bose-Einstein condensate into any Bloch state of an optical lattice (Fig.\ \ref{fig:scheme}).  First, a condensate is formed at $\mathbf{k}=0$ in the absence of an optical lattice.  Second, the condensate is accelerated to a momentum $\hbar \mathbf{k}$ lying in the $n$-th Brillouin zone.  Third, the lattice potential is  ramped on, mapping the condensate adiabatically into the $\mathbf{q} = \mathbf{k}$ Bloch state in the $n$-th band \cite{fall03lensing,brow05transport}.

We prepare Bose-Einstein condensates of $0.4-22\times10^{4}$ $^{87}$Rb atoms in an optical dipole trap with trap frequencies $\omega_{x,y,z} = 2 \pi \times (23,41, 46) $ Hz.  A vertically  ($\mathbf{z}$)  oriented light beam at 1064 nm wavelength, with its focus displaced from the condensate in the $x-y$ plane, is imposed for a variable time on the order of 1 ms.  The dipole force of this beam accelerates the condensate at 6 m/s$^2$ in the $\mathbf{y}$ direction.  The $1/e^2$ radius of the beam (85 $\mu$m) is larger than the $R_\mathrm{TF} \sim 10 \, \mu$m Thomas-Fermi radii of the condensate, reducing effects of the imposed dipole potential curvature on the gas.
\footnote{Calculations following  \cite{castin96trap} lead to an estimated increase in the peak density of $15\%$ due to the curvature of the imposed beam.}

After the accelerating optical potential is switched off, we gradually impose an optical kagome lattice in the horizontal plane \cite{jo12kag,thom17scaling}.  This lattice is formed by overlaying two triangular lattices, created with short- (SW, 532 nm) and also long-wavelength (LW, 1064 nm) light with in-plane polarization.  The depths of the two sublattices are increased to their final values in $T=1.2$ ms \footnote{The SW (LW) lattices are ramped up from initial values $V_i = h \times 0.5$ ($0.4$) kHz, respectively,  to their final values $V_f = V_{\text{SW}}$ ($V_{\text{LW}}$) as $V(t) = V_\text{i} + (V_\text{f}-V_\text{i})(e^{\alpha t/T}-1)/(e^\alpha-1)$, where $\alpha = 2.5$ and $T = 1.2$ ms.  The initial values, required for stabilization, do not affect the accelerating condensate.  Ramp parameters are chosen based on simulations that indicate a minimum state fidelity of 99\%, 95\% and 90\% for the $n=1$, $n=3$ and $n=4$ bands, respectively.}. The potential along the $\mathbf{z}$ axis is unmodified by the lattice beams and remains loosely confining. 

We characterize the lattice-bound condensate by measuring its momentum distribution.  
After allowing the condensate to evolve within the lattice for $\leq$ 350 $\mu$s, we suddenly switch off both the optical lattice and dipole trap. The atoms expand into a loosely confining magnetic trap, and then, after a quarter-cycle of harmonic oscillation, are imaged.  This technique maps a Bloch state into a reciprocal lattice of sharply peaked atomic distributions. 

Letting $q_K$ be the magnitude of the quasimomentum at the first $K$ point, accelerating atoms along $\mathbf{y}$ to a wavevector $k/q_K$ in the range of $(0, 1)$ places the lattice-trapped gas into the first band, of $(1, 1.5)$ into the third band, and of $(1.5, 2)$ into the fourth band of the kagome lattice (Fig.\ \ref{fig:scheme}(c)).  Since the condensate has negligible widths in momentum space approximated by $\hbar/R_{\text{TF}} = 0.02 (0.01) \hbar q_K$ in the $\mathbf{y}$($\mathbf{x}$) direction,  the entire quantum gas can be loaded into a single band as long as one avoids the edges of the Brillouin zone, where our adiabatic-loading scheme fails.

\begin{figure}[t]
\includegraphics{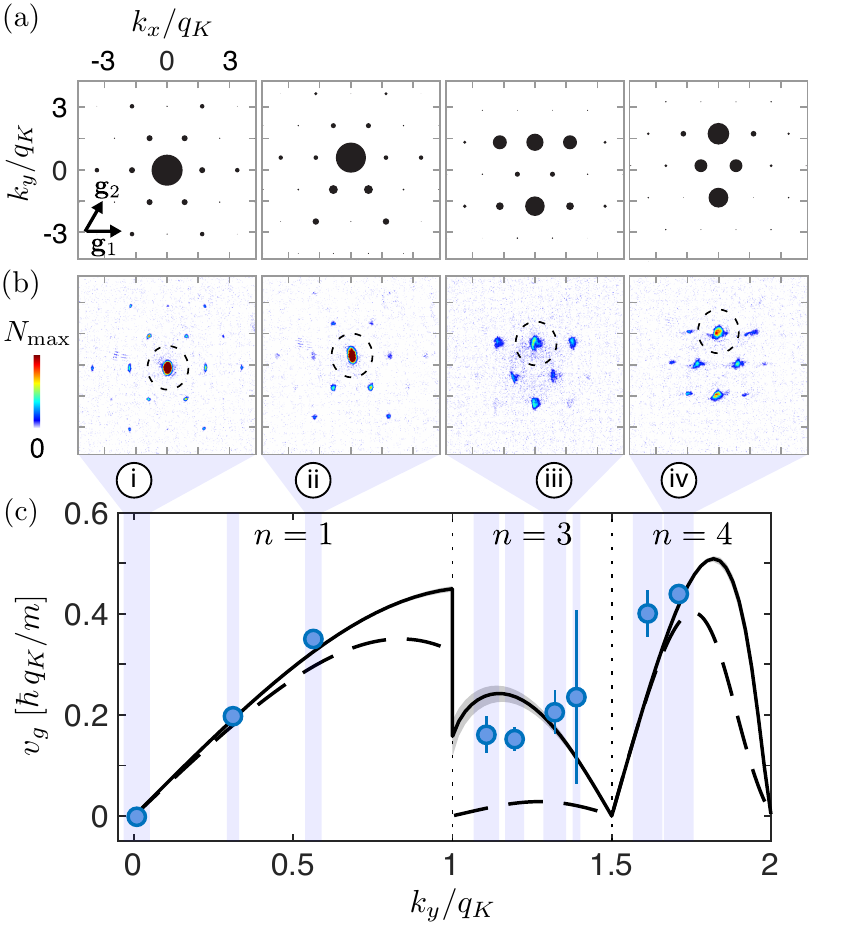}
\caption{Measured group velocity in the kagome lattice.  Momentum distributions are shown (a: non-interacting Bloch theory; b: measured) at four representative initial $\mathbf{k}$ (indicated by dashed circles in (b)) with labels corresponding to those in Fig.\ \ref{fig:scheme}(d).  Basis reciprocal lattice vectors $\mathbf{g}_{1,2}$ are shown.  (c) Measured $v_g = \mathbf{v}_g \cdot \mathbf{y}$.  The initial wavevector $\mathbf{k}$ is measured for each experimental repetition.  Data with $k_y$ within a binning range (blue shaded bars) of about 0.1 $q_K$ are averaged, with 3-8 measurements per bin.  Error bars are standard mean errors.  Data agree with calculations that include effects of interactions at a peak density of $n_0 = 5.4(5) \times10^{13} \text{cm}^{-3}$  (solid black curve, gray region indicates effect of density uncertainty), and disagree with non-interacting band-theory predictions (black dashed line). Final lattice depths are $(V_\text{SW},V_\text{LW}) = h\times (25, 15)$ kHz.}
\label{fig:GV}
\end{figure}

Representative momentum distributions at four points along this trajectory are shown in Fig.\ \ref{fig:GV}(b).  
Qualitatively, these distributions match with those calculated for non-interacting atoms in our kagome lattice (Fig.\ \ref{fig:GV}(a)).  This qualitative agreement indicates that our procedure places the entire population of the condensate into excited Bloch states, including into the $n=3$ band whose (non-interacting) band dispersion is near zero \footnote{Faster ramp-up of the lattice potential leads to lower Bloch-state preparation fidelity.  For such faster ramp-up, we observe temporal oscillations in the momentum-space populations at the final lattice depths, indicating that the atoms occupy a superposition of multiple bands.
In contrast, for the $T=1.2$ ms ramp-time data presented in this work, we observe no such oscillations.}.

However, a quantitative analysis reveals dramatic differences from the non-interacting band model. We focus on the group velocity $\mathbf{v}_g = \hbar^{-1} \nabla_\mathbf{q} E_{n}(\mathbf{q})$. By the Hellmann-Feynman theorem \cite{gutt32,hell33,feyn39forces}, which applies regardless of interaction strength, the group velocity is related to the mean velocity $\mathbf{v}_g = \langle \hbar \mathbf{k}/m\rangle$ of the Bloch state, with $m$ being the atomic mass.

Experimentally, we measure this mean velocity by applying spatial fits to the imaged distribution in a region surrounding each peak to determine the population  $N_\mathbf{G}$ of atoms in the $\hbar (\mathbf{q} + \mathbf{G})$ momentum states, where $\mathbf{G}$ are reciprocal lattice vectors.  We then take the weighted average $\mathbf{v}_g = (\hbar/m) (\sum_\mathbf{G} N_\mathbf{G} (\mathbf{q} + \mathbf{G})) / (\sum_\mathbf{G} N_\mathbf{G})$.  

The observed group velocity disagrees profoundly with the non-interacting band structure result.  In particular, whereas the non-interacting band theory predicts a near-zero group velocity within the $n=3$ kagome-lattice band, we observe a gas of atoms loaded into that band to have a significantly higher group velocity.

To explain this disgreement,
we consider effects of interatomic interactions.  Band structure may apply in interacting systems in different ways.  One approach is to consider weak excitations atop an interacting system at equilibrium and characterized by lattice symmetry. The spectrum of low-lying bands of excitations of interacting lattice-trapped quantum gases have been measured, for example, through optical Bragg scattering \cite{clem09bragg,du10bragg,erns10bragg,liu11bragg}.

Alternately, band structure may be used to describe the far from equilibrium state of a lattice-bound gas that is driven in its entirety into an excited Bloch state within the lattice.  For an interacting condensed gas, interactions are treated at the mean-field level by calculating the Bloch states of the lattice Gross-Pitaevskii equation,

\begin{eqnarray}
	\left(-\frac{\hbar^2}{2m}\nabla^2 + V(\mathbf{r}) + g |\Psi(\mathbf{r})|^2\right)\Psi(\mathbf{r})  = E_n(\mathbf{q}) \Psi(\mathbf{r})
\label{eq:GPE}
\end{eqnarray}
where $V(\mathbf{r})$ is the optical kagome lattice potential, $g = 4\pi\hbar^2a/m$ with $a$ being the s-wave scattering length, and $\Psi(\mathbf{r})$ is the condensate wavefunction.  Self-consistent Bloch-state solutions are found where the density $|\Psi(\mathbf{r})|^2$ is symmetric under translation by lattice vectors. The state $\Psi(\mathbf{r})$ is then a Bloch state for a non-interacting gas in an overall potential that is the sum of the applied lattice potential $V(\mathbf{r})$ and the spatially periodic interaction energy $g_0 |\Psi(\mathbf{r})|^2$.  Such Bloch-state solutions have been studied in the context of quantum gases,  highlighting the emergence of additional solutions beyond those in the non-interacting case (swallowtails and states with doubled spatial period), of related dynamical phenomena such as nonlinear Landau-Zener tunneling and hyteresis 
\cite{wu00landauzener,liu02landauzener,wu02bloch,diako02loop,muel02hysteresis,machh03,seam05nonlinear,chen11fingerprint, koll16loop, machholm04doubling}, and modulational instability \cite{wu01instability, konotop02instability,04fallaniinstability}.  Nonlinear Bloch modes also arise in nonlinear photonic crystals \cite{trag06bloch} and exciton-polariton condensates \cite{ches16excitonpolariton}.

The group velocity determined from Eq.\ \ref{eq:GPE} agrees with our measurements (Fig.\ \ref{fig:GV}(c)) \footnote{For this calculation, we identify those solutions of Eq.\ \ref{eq:GPE} that map smoothly onto the Bloch state of the non-interacting system with the corresponding band index.}.  
The lattice Gross-Pitaevskii equation also predicts specific differences in the Bloch-state momentum distribution between the non-interacting and interacting cases. Such differences are evident in our data, although not as robustly as the overall change in the group velocity.
Both numerical and experimental data show that the effect of interactions on the group velocity is most pronounced in the $n=3$ band, adding significant dispersion to a band that, in the non-interacting case, is nearly flat.

\begin{figure}[t]
\includegraphics{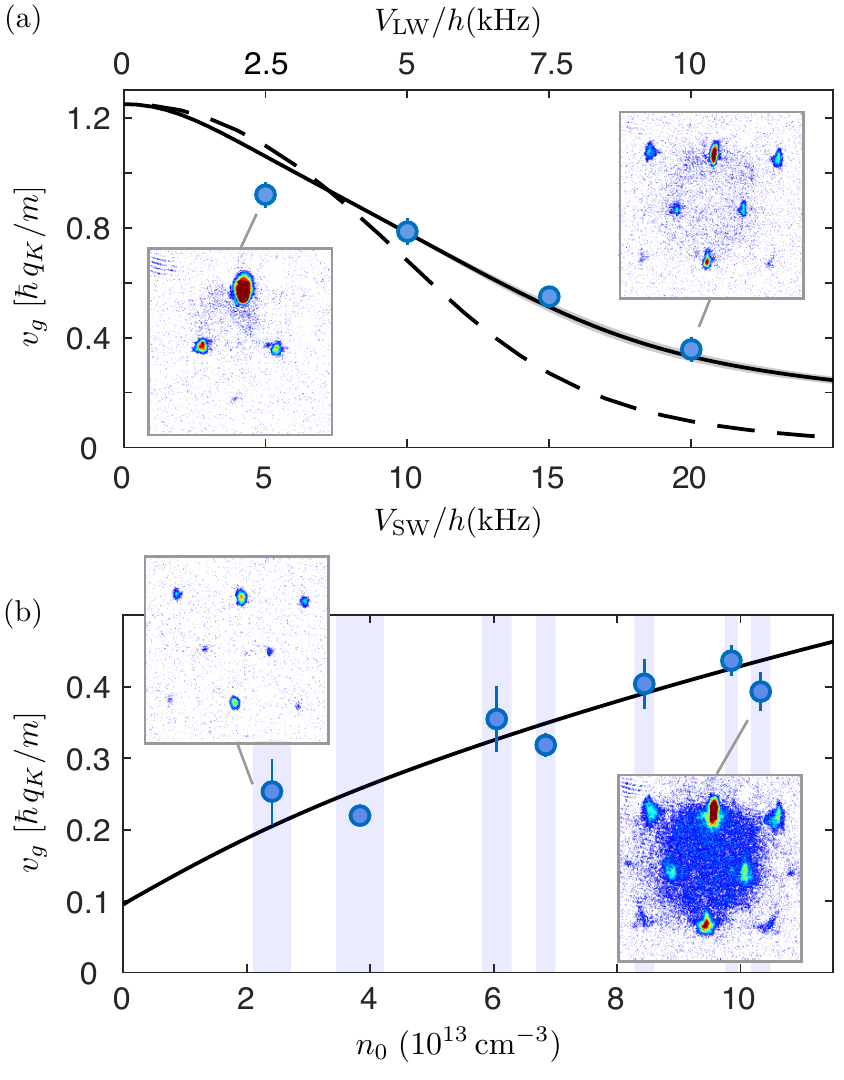}
\caption{Dependence of $v_g$ on lattice depths and densities measured at a fixed initial wavevector of $k_y = 1.25\, q_K$. (a) $v_g$ measured with a peak density $n_0 = 6.2(6) \times 10^{13} \text{cm}^{-3}$ and different lattice depths, with $V_{\text{SW}}/V_{\text{LW}}$ kept at 2. While band theory (dashed black curve) predicts that $v_g$ is suppressed quickly at increasing lattice depths, data (each point represents the average of 4-7 measurements) show a significantly smaller rate of suppression, in agreement with Gross-Pitaevskii equation predictions (solid black curve, gray region indicates effects of density uncertainty). (b) $v_g$ measured at fixed lattice depths $(V_{\text{SW}},V_{\text{LW}}) = h\times(20,10)$ kHz increases monotonically with number density as predicted by the Gross-Pitaevskii equation (black solid curve), clearly indicating interaction effects. Data within a small binning range (blue shaded bars) of densities, which are determined up to $10\%$ systematic uncertainty, are averaged, with between 3-8 measurements per bin. Error bars are standard mean errors. Insets are single-shot images taken at indicated settings.}
\label{fig:depthscan}
\end{figure}

We characterize the interaction-induced  distortion of the flat band of the kagome band structure by two additional experiments.  We focus on the initial wavevector $k_y = 1.25\, q_K$, which lies in the middle of the portion of the $n=3$ Brillouin zone accessed by our procedure.  In one experiment (Fig.\ \ref{fig:depthscan}(a)), we measure  $v_g$ for identically prepared condensates that are loaded into lattices of increasing depth. 
For the non-interacting case, the calculated group velocity tends quickly to zero as the lattice is deepened and approaches the kagome-geometry, tight-binding limit.  In contrast, in the presence of interactions, while $v_g$ diminishes for increasing lattice depth, it does so only slowly and lies significantly higher than the non-interacting result.

In a second experiment, we study the dependence of $v_g$ on the interaction strength by varying the density of the gas.  We load condensates within initial peak densities $n_0$ in the range of $2 - 11 \times 10^{13}$ cm$^{-3}$ into the aforementioned Bloch state of the lattice, and find $v_g$ increases with gas density (Fig.\ \ref{fig:depthscan}(b)), showing that the flat band of the kagome lattice acquires a dispersion that increases with interaction energy.

\begin{figure}[t]
		\includegraphics{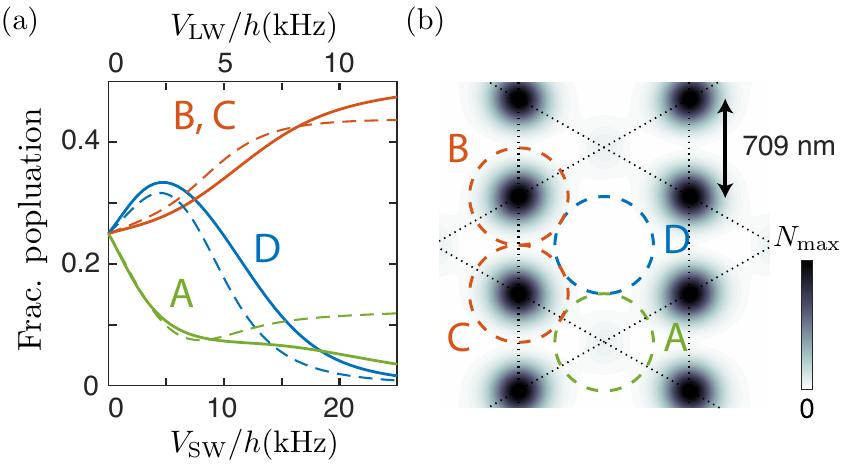}
	\caption{Real-space distribution of atoms in the $n=3$, $k_y=1.25\, q_K$ Bloch state. (a) Calculated fractional population in the four kagome-lattice sites in a unit cell as a function of lattice depths with $V_{\text{SW}}/V_{\text{LW}} = 2$ (dashed line: band theory; solid line: Gross-Pitaevskii equation).  (b) Real-space distribution of atoms in a lattice at $(V_{\text{SW}}, V_{\text{LW}}) = h\times (20, 10)$ kHz calculated by solving the Gross-Pitaevskii  equation with $n_0 = 6.2\times10^{13}\text{cm}^{-3}$}
	\label{fig:realspace}
\end{figure}

The picture that emerges from our findings is that the interaction energy of atoms within a Bloch state adds to and distorts the lattice potential, so that the transport properties of atoms in the resulting overall lattice potential differ dramatically from those dictated by the optical lattice on its own. This distortion is evident in the real-space atomic distribution predicted by the lattice Gross-Pitaevskii equation.  In Fig.\ \ref{fig:realspace}, we consider again the $n=3$, $k_y = 1.25 \, q_K$ Bloch state, and calculate the population fractions in the four sites of the lattice unit cell, with A, B and C being the three allowed sites in the kagome lattice, and D being the site excluded from the lattice as $V_\mathrm{LW}$ is increased.  In the deep lattice, the population becomes concentrated largely in just two sites of the kagome lattice (B and C).  The unequal population of atoms in the kagome lattice sites leads to a mean-field interaction potential that departs from the kagome-lattice geometry.  It is then not surprising that the band-structure of this distorted overall lattice potential no longer supports a flat $n=3$ band.

The quenching of kinetic energy amplifies the effects of interactions on a many-body system that occupies a non-dispersing band.  For interacting bosons equilibrating within the flat band of the $J<0$ kagome lattice, several low-temperature states have been discussed, including Wigner crystal and supersolid phases \cite{hube10flat}, a superfluid residing at the band minima of an interaction-renormalized energy band \cite{you12kagome}, and fractional-filling Mott insulator states \cite{para13}.  Our demonstrated ability to place a $^{87}$Rb gas into the $n=3$ band of a $J>0$ kagome lattice raises the possibility that such predictions could be tested in the non-equilibrium setting of bosons evolving transiently within an excited band.

However, we observe the excited-band populations in our experiment to be unstable to decay.  Such decay is seen, for example, in the momentum distribution of the highest-density gas shown in Fig.\ \ref{fig:depthscan}(b), where the sharp momentum peaks of the coherent Bloch state give way to a broad momentum distribution.  This broad distribution grows to a significant fraction of the total atom population within hundreds of $\mu$s, with shorter lifetimes seen for higher-density gases in higher-depth lattices.  Through band mapping, we determine that this decay produces atoms predominantly in the ground band.  Further, examining also the decay of atoms prepared in the $n=2$ (reached by accelerating the gas initially along $\mathbf{x}$) or $n=4$ bands, we observe the $n=3$ Bloch state to decay most rapidly.  It remains to be seen whether atoms prepared in energy extrema of the renormalized $n=3$ band, predicted to lie at the $\Gamma$ and $K$ points, show greater stability.  The decay to lower bands occur through collisions that transfer band energy into the loosely confined $\mathbf{z}$ direction of motion.  It may be possible to forestall such decay by adding an additional confining lattice along $\mathbf{z}$.

In conclusion, in searching for experimental evidence of the non-dispersing nature of the $n=3$ band of the kagome lattice, we find, instead, that interactions among atoms placed within that band lead to significant band-structure renormalization.  The interaction-based distortion of the band structure is seen by directly measuring the group velocity of the Bloch state, and finding it to be significantly larger than predicted by non-interacting band theory. The emergence of a modified overall lattice structure generated by atoms within a lattice is reminiscent of experiments on quantum gases within optical cavities \cite{gopa09emergent,mott12roton}.  In the optical-cavity experiments, the emergent lattice is generated by  light-induced extended-range atomic interactions, whereas, in the present work, the emergent potential is produced by direct local interactions.  Future work may examine which aspects of the non-interacting band structure become invalid for lattice-trapped interacting systems, e.g.\ by studying the renormalization of band gaps and the interplay between hystersis and band-geometry effects.

We thank Storm Weiner and Leon Lu for experimental assistance and valuable discussions.
This work was supported by the NSF, and by the AFOSR and ARO through the MURI program (grant numbers FA9550-14-1-0035 and W911NF-17-1-0323, respectively). Govind Unnikrishnan acknowledges support by the Austrian Science Fund FWF within the DK-ALM (W1259-N27).

\bibliographystyle{apsrev4-1}
\bibliography{Manuscript_final.bbl}

\end{document}